# Integrating Zero-Shot Classification to Advance Long COVID Literature: A Systematic Social Media–Centered Review

Nirmalya Thakur

Department of Electrical Engineering and Computer Science, South Dakota School of Mines and Technology, Rapid City, SD 57701, USA
nirmalya.thakur@sdsmt.edu

**Abstract.** Long COVID continues to challenge public health by affecting a significant segment of individuals who have recovered from acute SARS-CoV-2 infection yet endure prolonged and often debilitating symptoms. Social media has emerged as a vital resource for those seeking real-time information, peer support, and validating their health concerns related to Long COVID. This paper examines recent works focusing on mining, analyzing, and interpreting user-generated content on social media platforms such as X (formerly Twitter), Reddit, Facebook, and YouTube to capture the broader discourse on persistent post-COVID conditions. A novel transformer-based zero-shot learning approach serves as the foundation for classifying research papers in this area into four primary categories: Clinical or Symptom Characterization, Advanced NLP or Computational Methods, Policy, Advocacy, or Public Health Communication, and Online Communities and Social Support. This methodology showcases the adaptability of advanced language models in categorizing research papers without predefined training labels, thus enabling a more rapid and scalable assessment of existing literature. This review highlights the multifaceted nature of Long COVID research, where computational techniques applied to social media data reveal insights into narratives of individuals suffering from Long COVID. This review also demonstrates the capacity of social media analytics to inform clinical practice and contribute to policy making related to Long COVID.

**Keywords:** Long COVID, COVID-19, Zero-Shot Learning, social media, Twitter, Reddit, Facebook, and YouTube

## 1   Introduction

In December 2019, an outbreak of coronavirus disease 2019 (COVID-19), caused by the severe acute respiratory syndrome coronavirus 2 (SARS-CoV-2), began in China. [1,2]. Even though SARS-CoV-2 is similar in origin to SARS-CoV and MERS-CoV, it has affected public health globally at a much greater scale than any prior coronavirus outbreaks [3]. Early containment efforts, including measures by the Chinese government, did not prevent the disease from rapidly crossing regional and international boundaries [4], leading the World Health Organization (WHO) to declare COVID-19 a global pandemic on March 11, 2021 [5]. According to the WHO, confirmed cases were 776,841,264 worldwide, with 7,075,468 reported deaths as of 10 November 2024 [6].

Although many individuals recover from the acute infection caused by SARS-CoV-2, a significant subset experiences symptoms that remain or appear after what might have been presumed clinical recovery. This phenomenon, known as Long COVID, has been described since the earliest days of the pandemic to include persistent or emerging physical and psychological challenges [7-9]. As per [10], *"Long COVID is defined as a chronic condition that occurs after SARS-CoV-2 infection and is present for at least 3 months. Long COVID includes a wide range of symptoms or conditions that may improve, worsen, or be ongoing"*.

Individuals who experience Long COVID commonly face a broad set of symptoms that may disrupt daily routines and overall well-being. Frequently reported symptoms of Long COVID include shortness of breath, cough, persistent fatigue, post-exertional malaise, difficulty concentrating, memory changes, recurring headache, lightheadedness, fast heart rate, sleep disturbance, problems with taste or smell, bloating, constipation, and diarrhea [11,12]. In more complex scenarios, patients are diagnosed with interstitial lung disease and hypoxemia, cardiovascular disease and arrhythmias, cognitive impairment, mood disorders, anxiety, migraine, stroke, blood clots, chronic kidney disease, postural orthostatic tachycardia syndrome (POTS) and other forms of dysautonomia, myalgic encephalomyelitis/chronic fatigue syndrome (ME/CFS), mast cell activation syndrome (MCAS), bromyalgia, connective tissue diseases, hyperlipidemia, diabetes, and autoimmune disorders such as lupus, rheumatoid arthritis, and Sjogren's syndrome [11,13-15].

Such symptoms may persist for three months beyond the initial SARS-CoV-2 infection or even exceed a year [16]. Although some individuals gradually improve, others experience lingering or fluctuating complications that may profoundly affect their physical, psychological, and social health [17,18]. Most treatment approaches revolve around consistent monitoring and symptom-specific care. Clinicians commonly reference established guidelines when managing symptoms of Long COVID and any accompanying conditions, such as diabetes, high blood pressure, or POTS, to reduce future complications and enhance the patient's quality of life [19]. Many patients benefit from a combination of therapies - ranging from medications targeting pain or sleep challenges to physical or occupational rehabilitation - along with psychological support to manage both the physical and emotional aspects of Long COVID [20,21]. In addition to this, even though certain medications such as paracetamol or NSAIDs appear to help with specific Long COVID symptoms like fever, there is still no standardized treatment to address the entire spectrum of Long COVID symptoms [22-25].

Social media has been a critical venue for public discussion of COVID-19 since its initial cases in December 2019, evolving into a resource for people seeking real-time information and community support [26-31]. As this pandemic advanced, platforms such as X (formerly Twitter) [32,33], TikTok [34,35], Instagram [36,37], Facebook [38,39], YouTube [40,41], Reddit [42,43], LinkedIn [44,45], Clubhouse [46,47], Discord [48,49], and Snapchat [50,51], became pivotal for gathering firsthand insights into ongoing patient experiences. Traditional methods like surveys and interviews can be constrained by time and location, whereas social media allows continuous, unfiltered accounts of Long COVID manifestations and daily struggles. Individuals suffering from Long COVID can document their symptoms, exchange practical advice, and discuss personal setbacks or milestones on social media, leading to the generation of Big

Data that researchers from different disciplines and healthcare professionals may analyze to identify evolving patterns.

As Long COVID presents multifaceted medical, social, and emotional issues, there has been growing interest in leveraging online platforms to study it from multiple angles. Social media channels facilitate global conversations that can reveal differences in experiences related to healthcare access, post-infection complications, or even public awareness of the severity of a health-related condition [52-54]. Over time, these virtual spaces have also fostered advocacy and grassroots efforts. Hashtags like #LongCOVID [55] have given patients and advocates an active role in discussing everything from specialized clinics to mental health support [56]. Observations of this activity underscore how large-scale social media data can shape public health discourse [57] and even influence policies [58] addressing health-related conditions that are often misunderstood or underdiagnosed.

A review paper that categorizes the existing work in this domain is expected to play a crucial role in advancing knowledge. Studies on Long COVID and social media vary widely, incorporating sentiment analysis, network analysis, qualitative content studies, and more. Combining or comparing such research can clarify where the field has gathered robust evidence, where it lacks conclusive data, and which areas still need systematic exploration. Recent works have drawn on patient narratives to refine clinical definitions or inform the development of Long COVID care frameworks, yet there is a crucial need to consolidate these findings. Examining studies in this area under broader thematic groupings is expected to highlight the progress made so far and pinpoint unresolved questions, methodological gaps, and ethical considerations surrounding patient data. Addressing this research gap serves as the main motivator for this review.

This review focuses on four broad areas - Clinical or Symptom Characterization, Advanced NLP or Computational Methods, Policy, Advocacy, or Public Health Communication, Online Communities and Social Support, in the context of Long COVID-related research works that specifically focus on mining, analyzing, and interpreting social media data. It is relevant to mention that this review not only investigates the broad spectrum of research on Long COVID across various social media platforms but also integrates a novel zero-shot classification pipeline that organizes recent works in this field into distinct categories. This classification process was performed without explicit task-specific training using a transformer-based model configured for zero-shot learning. This dual perspective highlights both a comprehensive review of the literature and a demonstration of how an advanced language model can streamline the analysis of research papers on Long COVID. This review also discusses current research limitations and proposes future work directions that are expected to benefit both scientific communities and those living with Long COVID.

Although these studies appear classified into distinct areas in this review paper, many works address multiple facets of Long COVID research, rendering any classification flexible rather than absolute. For instance, a paper listed under "Online Communities and Social Support" may also perform a detailed sentiment analysis that aligns with "Advanced NLP or Computational Methods". Such overlaps arise naturally in interdisciplinary research, especially when varied computational methods - like sentiment analysis, topic modeling, and network analysis - are applied to the extensive social media discussions surrounding patient experiences, advocacy efforts, and policy implications. The four broad areas presented here are an organizational guide, highlighting a primary thematic focus without dismissing other significant aspects of each study.

The rest of this paper is organized as follows. First, the methodology is presented, describing the search strategy, inclusion criteria, and the steps taken to apply a zero-shot learning model. Then, the results from this automated classification process are discussed, followed by a detailed examination of each study identified in the review. Research gaps and directions for future work are then explored, emphasizing how interdisciplinary approaches and advanced computational tools might enrich the current understanding of Long COVID. Finally, the paper summarizes key findings and highlights the potential impact of integrating social media analytics into ongoing research on persistent COVID-19 symptoms.

## 2  Methodology

A broad literature search was carried out across multiple scholarly databases, including PubMed, Scopus, Web of Science, and Google Scholar, to identify studies focused on mining, analyzing, and interpreting the public discourse about Long COVID on social media. This search aimed to capture Long COVID-related research across diverse fields, such as computer science, health sciences, and social sciences. No papers published before 2020 were included, as the COVID-19 outbreak began in December 2019. The search terms used included - "Long COVID," "post-COVID," "chronic COVID" - as well as keywords indicative of social media use, such as "Twitter", "TikTok", "Instagram", "Facebook", "YouTube", "Reddit", "LinkedIn", "Clubhouse", "Discord", and "Snapchat". In addition, terms like "sentiment analysis," "topic modeling," and "network analysis" were included to ensure the retrieval of studies that used computational or statistical methods to examine the public discourse on these platforms. By blending health-related terminology with references to digital platforms and relevant analytical techniques, the search strategy was designed to capture the full breadth of scholarly work investigating individuals' ongoing experiences with Long COVID.

Studies were selected for inclusion if they used social media data to investigate any aspect of Long COVID. The main inclusion criteria were that the articles utilized a recognized research methodology - whether qualitative, quantitative, or mixed methods and analyzed data gathered primarily from social media platforms. The selection process also considered whether the authors had sufficiently detailed the nature of their quantitative or qualitative approach. Moreover, ethical practices regarding user data, such as anonymization or compliance with platform terms of service, were taken into account to ensure that privacy concerns were handled responsibly. Studies that just mentioned Long COVID were excluded. Research works such as editorials, letters to the editor, or general news articles, which usually lack methodological details, were also excluded. If the essential aspects of a paper were missing - for example, neglecting to report how data were collected - those were also removed from consideration. This approach aimed to retain a set of methodologically sound articles that offered substantive insights into the public discourse about Long COVID on social media platforms.

Upon applying these inclusion and exclusion criteria, a total of 40 studies were selected for this review. These works represented a range of methods, including sentiment analysis, qualitative content analysis, topic modeling, and network analysis, and they addressed multiple social media platforms, such as X (formerly Twitter), Reddit, and Facebook. Thereafter, a transformer-based zero-shot classification model was developed to classify these papers into one out of the four thematic categories described below. Although a few papers fit under multiple themes, each was placed wherever its primary emphasis appeared strongest.

(i) Clinical or Symptom Characterization ("Symptom Characterization"): Research that primarily aims to identify, list, or quantify the variety of Long COVID symptoms, usually from social media data. The studies may include statistical analysis but do not necessarily perform extensive sentiment or topic modeling. Their main motivation is to gather clinical or epidemiological insights from user posts.

(ii) Advanced NLP or Computational Methods ("NLP and Modeling"): Studies that specifically emphasize methods like deep transformer networks, topic modeling, sentiment analysis, and other elaborate computational approaches. This goes beyond a simple symptom count; it highlights a methods-heavy lens on analyzing data.

(iii) Policy, Advocacy, or Public Health Communication ("Policy and Advocacy"): Papers exploring how organizations, governments, or communities develop health communications, handle policy issues, and communicate guidelines.

(iv) Online Communities & Social Support ("Community and Support"): Studies focusing on how individuals find emotional or experiential support on social media, the way they exchange personal stories, or how group dynamics form around shared experiences. The main emphasis is on the psychosocial aspect, and the support social media platforms provide.

Thereafter, a transformer-based zero-shot classification model was developed, which was set up to assign the 40 papers to these categories. This model did not require any training using any labeled dataset. The process by which this model worked is described below:

Formally, let "text" be a study's abstract, and suppose we have candidate labels {$c_1$, $c_2$, …, $c_N$}. The model used a scoring function to determine the alignment between "text" and a label $c_k$. Equation (1) shows how the best-matching category, denoted as ĉ(text) was determined.

$$\hat{c}(\text{text}) = \operatorname*{argmax}_{k} p_k(\text{text}) \qquad (1)$$

In Equation (1), $p_k(\text{text})$ represents the probability that text belongs to category $c_k$. In practical terms, the system prompted the model with textual descriptions of each category and the study's abstract. It then computed a scalar score $s(\text{text}, k)$ (as shown in Equation (2)) that measured how well text matched the meaning or intent of label $c_k$. A softmax function then normalized these raw scores, producing probabilities for all categories as shown in Equation (3).

$$s(\text{text}, k) = f\Big(\text{text}, \text{prompt}(k)\Big) \qquad (2)$$

$$p_k(\text{text}) = \frac{\exp\left[s(\text{text}, k)\right]}{\sum_{j=1}^{N} \exp\left[s(\text{text}, j)\right]} \tag{3}$$

In essence, whichever category attained the largest probability was selected as the label by the model for that document. This methodology is called zero-shot learning [59-61] because the model does not require an example corpus manually labeled under these same categories. Instead, the model draws from its vast, pre-trained language representations to infer whether a given textual data aligns more with, for instance, a "Symptom Characterization" theme or a "Community and Support" theme. A program was written in Python 3.10 to develop and implement this transformer-based zero-shot classification model. The results of the same are presented and discussed in Section 3.

## 3   Results of Zero-Shot Classification

This section presents the results of the transformer-based zero-shot classification model applied to the 40 research papers [62-102] that met the inclusion criteria of this review. The underlying premise of zero-shot learning is that when prompted with suitable descriptors, a well-trained language model can identify the most relevant label for textual data, even if the model has never seen concrete examples corresponding to that label during training. This eliminates the need for time-intensive and resource-heavy data labeling processes, which is especially advantageous in emerging research areas such as COVID-19-related research where existing taxonomies may be incomplete or still evolving [103-106]. In addition to the inherent benefit of not requiring pre-labeled data, this methodology also provided a structured and transparent way to allocate these 40 research papers into distinct categories. By integrating zero-shot learning with carefully curated dictionary-based keyword matching, it became possible to identify and highlight the thematic focus of each study and classify it into one of the four categories - Clinical or Symptom Characterization, Advanced NLP or Computational Methods, Policy, Advocacy, or Public Health Communication, Online Communities and Social Support. The dictionary-based scores helped complement the probabilistic outputs from the zero-shot classification model, thereby refining the final assessments of each study's thematic focus. This synergy proved particularly useful for works where technical and clinical terminologies might intersect, making it difficult to rely solely on either semantic features or explicit keyword usage. The result was a more robust and interpretable categorization pipeline that could be applied to other domains as well with minimal customization effort. Table 1 shows the results where the author list, title, and classification label of each paper are shown.

Table 1: Results of applying zero-shot learning to classify the 40 papers that met the inclusion criteria

| Full Author List | Title | Classification Label |
|---|---|---|
| Yu-Bo Fu [62] | Investigating public perceptions regarding the Long COVID on Twitter using sentiment analysis and topic modeling | NLP and Modeling |

| Alex Rushforth, Emma Ladds, Sietse Wieringa, Sharon Taylor, Laiba Husain and Trisha Greenhalgh [63] | Long Covid – The illness narratives | Policy and Advocacy |
|---|---|---|
| David Russell, Naomi J. Spence, Jo-Ana D. Chase, Tatum Schwartz, Christa M. Tumminello and Erin Bouldin [64] | Support amid uncertainty: Long COVID illness experiences and the role of online communities | Community and Support |
| Francesco Meledandri [65] | The Impact of Polarised Social Media Networking Communications in the #Longcovid Debate between Ideologies and Scientific Facts | Community and Support |
| Shubh Mohan Singh and Chaitanya Reddy [66] | An Analysis of Self-reported Longcovid Symptoms on Twitter | Symptom Characterization |
| Nida Ziauddeen, Deepti Gurdasani, Margaret E O'Hara, Claire Hastie, Paul Roderick, Guiqing Yao and Nisreen A Alwan [67] | Characteristics of Long Covid: findings from a social media survey | Symptom Characterization |
| Abeed Sarker and Yao Ge [68] | Long COVID symptoms from Reddit: Characterizing post-COVID syndrome from patient reports | Symptom Characterization |
| Juan M. Banda, Nicola Adderley, Waheed-Ul-Rahman Ahmed, Heba AlGhoul, Osaid Alser, Muath Alser, Carlos Areia, Mikail Cogenur, Krisitina Fišter, Saurabh Gombar, Vojtech Huser, Jitendra Jonnagaddala, Lana YH Lai, Angela Leis, Lourdes Mateu, Miguel Angel Mayer, Evan Minty, Daniel Morales, Karthik Natarajan, Roger Paredes, Vyjeyanthi S. Periyakoil, Albert Prats-Uribe, Elsie G. Ross, Gurdas Singh, Vignesh Subbian, Arani Vivekanantham and Daniel Prieto-Alhambra [69] | Characterization of long-term patient-reported symptoms of COVID-19: an analysis of social media data | Symptom Characterization |
| Daisy Massey, Diana Berrent and Harlan Krumholz [70] | Breakthrough Symptomatic COVID-19 Infections Leading to Long Covid: Report from Long Covid Facebook Group Poll | Symptom Characterization |
| Sam Martin, Macarena Chepo, Noémie Déom, Ahmad Firas Khalid and Cecilia Vindrola-Padros [71] | "#LongCOVID affects children too": A Twitter analysis of healthcare workers' sentiment and discourse about Long COVID in children and young people in the UK | Symptom Characterization |

| Authors | Title | Category |
|---|---|---|
| Elham Dolatabadi, Diana Moyano, Michael Bales, Sofija Spasojevic, Rohan Bhambhoria, Junaid Bhatti, Shyamolima Debnath, Nicholas Hoell, Xin Li, Celine Leng, Sasha Nanda, Jad Saab, Esmat Sahak, Fanny Sie, Sara Uppal, Nirma Khatri Vadlamudi, Antoaneta Vladimirova, Artur Yakimovich, Xiaoxue Yang, Sedef Akinli Kocak and Angela M. Cheung [72] | Using Social Media to Help Understand Long COVID Patient Reported Health Outcomes: A Natural Language Processing Approach | Symptom Characterization |
| Lin Miao, Mark Last and Marina Litvak [73] | An Interactive Analysis of User-reported Long COVID Symptoms using Twitter Data | Symptom Characterization |
| Guocheng Feng, Huaiyu Cai and Wei Quan [74] | Exploring the Emotional and Mental Well-Being of Individuals with Long COVID Through Twitter Analysis | Symptom Characterization |
| Alexis Jordan and Albert Park [75] | Understanding the Long Haulers of COVID-19: Mixed Methods Analysis of YouTube Content | NLP and Modeling |
| Ikhwan Yuda Kusuma and Suherman Suherman [76] | The Pulse of Long COVID on Twitter: A Social Network Analysis | NLP and Modeling |
| Nirmalya Thakur [77] | Investigating and Analyzing Self-Reporting of Long COVID on Twitter: Findings from Sentiment Analysis | NLP and Modeling |
| Toluwalase Awoyemi, Ujunwa Ebili, Abiola Olusanya, Kayode E. Ogunniyi and Adedolapo V. Adejumo [78] | Twitter Sentiment Analysis of Long COVID Syndrome | Symptom Characterization |
| Sam Rhodehamel [79] | Digital Long Hauler Lifelines: Understanding How People with Long Covid Build Community on Reddit | Community and Support |
| Arinjita Bhattacharyya, Anand Seth and Shesh Rai [80] | The Effects of Long COVID-19, Its Severity, and the Need for Immediate Attention: Analysis of Clinical Trials and Twitter Data | Policy and Advocacy |
| Surani Matharaarachchi, Mike Domaratzki, Alan Katz and Saman Muthukumarana [81] | Discovering Long COVID Symptom Patterns: Association Rule Mining and Sentiment Analysis in Social Media Tweets | Symptom Characterization |
| Jonathan Koss and Sabine Bohnet-Joschko [82] | Social Media Mining of Long-COVID Self-Medication Reported by Reddit Users: Feasibility Study to Support Drug Repurposing | Symptom Characterization |
| Hanin Ayadi, Charline Bour, Aurélie Fischer, Mohammad Ghoniem and Guy Fagherazzi [83] | The Long COVID Experience from a Patient's Perspective: A Clustering Analysis of 27,216 Reddit Posts | Symptom Characterization |
| Camryn Garrett, Atefeh Aghaei, Abhishek Aggarwal and Shan Qiao [84] | The Role of Social Media in the Experiences of COVID-19 Among Long-Hauler Women: Qualitative Study | Community and Support |

| Authors | Title | Category |
|---|---|---|
| Linnea I. Laestadius, Jeanine P. D. Guidry, Andrea Bishop and Celeste Campos-Castillo [85] | State Health Department Communication about Long COVID in the United States on Facebook: Risks, Prevention, and Support | Policy and Advocacy |
| Juan S. Izquierdo-Condoy, Raul Fernandez-Naranjo, Eduardo Vasconez-González, Simone Cordovez, Andrea Tello-De-la-Torre, Clara Paz, Karen Delgado-Moreira, Sarah Carrington, Ginés Viscor and Esteban Ortiz-Prado [86] | Long COVID at Different Altitudes: A Country-wide Epidemiological Analysis | Symptom Characterization |
| Sara Santarossa, Ashley Rapp, Saily Sardinas, Janine Hussein, Alex Ramirez, Andrea E Cassidy-Bushrow, Philip Cheng and Eunice Yu [87] | Understanding the #longCOVID and #longhaulers Conversation on Twitter: Multimethod Study | Community and Support |
| Amélia Déguilhem, Joelle Malaab, Manissa Talmatkadi, Simon Renner, Pierre Foulquié, Guy Fagherazzi, Paul Loussikian, Tom Marty, Adel Mebarki, Nathalie Texier and Stephane Schuck [88] | Identifying Profiles and Symptoms of Patients With Long COVID in France: Data Mining Infodemiology Study Based on Social Media | Symptom Characterization |
| Elham Dolatabadi, Diana Moyano, Michael Bales, Sofija Spasojevic, Rohan Bhambhoria, Junaid Bhatti, Shyamolima Debnath, Nicholas Hoell, Xin Li, Celine Leng, Sasha Nanda, Jad Saab, Esmat Sahak, Fanny Sie, Sara Uppal, Nirma Khatri Vadlamudi, Antoaneta Vladimirova, Artur Yakimovich, Xiaoxue Yang, Sedef Akinli Kocak and Angela M. Cheung [89] | Using Social Media to Help Understand Patient-Reported Health Outcomes of Post–COVID-19 Condition: Natural Language Processing Approach | Symptom Characterization |
| Nida Ziauddeen, Deepti Gurdasani, Margaret E. O'Hara, Claire Hastie, Paul Roderick, Guiqing Yao and Nisreen A. Alwan [90] | Characteristics and Impact of Long Covid: Findings from an Online Survey | Symptom Characterization |
| Ludovica Segneri, Nandor Babina, Teresa Hammerschmidt, Andrea Fronzetti Colladon and Peter A. Gloor [91] | Too Much Focus on Your Health Might Be Bad for Your Health: Reddit User's Communication Style Predicts Their Long COVID Likelihood | Symptom Characterization |

| Sai C. Reddy, Sanjana Kathiravan and Shubh M. Singh [92] | An Analysis of Self-reported Long COVID-19 Symptoms on Twitter | Symptom Characterization |
|---|---|---|
| Abeed Sarker [93] | Mining Long-COVID Symptoms from Reddit: What We Know So Far | Symptom Characterization |
| Esperanza Miyake and Sam Martin [94] | Long COVID: Online Patient Narratives, Public Health Communication, and Vaccine Hesitancy | Community and Support |
| Abeed Sarker and Yao Ge [95] | Mining Long-COVID Symptoms from Reddit: Characterizing Post-COVID Syndrome from Patient Reports | Symptom Characterization |
| Alexis Jordan and Albert Park [96] | Understanding the Plight of COVID-19 Long Haulers Through Computational Analysis of YouTube Content | NLP and Modeling |
| Brigitte Juanals and Jean-Luc Minel [97] | Using topic modeling and NLP tools for analyzing long Covid coverage by French press and Twitter | Community and Support |
| Erkan Ozduran and Sibel Büyükçoban [98] | A Content Analysis of the Reliability and Quality of YouTube Videos as a Source of Information on Health-Related Post-COVID Pain | Community and Support |
| Noémie Déom, Ahmad Firas Khalid, Sam Martin, Macarena Chepo, and Cecilia Vindrola-Padros [99] | Unlocking the Mysteries of Long COVID in Children and Young People: Insights from a Policy Review and Social Media Analysis in the UK | Policy and Advocacy |
| Erin T. Jacques, Corey H. Basch, Eunsun Park, Betty Kollia and Emma Barry [100] | Long Haul COVID-19 Videos on YouTube: Implications for Health Communication | Symptom Characterization |
| William David Strain, Ondine Sherwood, Amitava Banerjee, Vicky Van der Togt, Lyth Hishmeh and Jeremy Rossman [101] | The Impact of COVID Vaccination on Symptoms of Long COVID: An International Survey of People with Lived Experience of Long COVID | Symptom Characterization |
| Krittiya Wongtavavimarn [102] | Social Support and Narrative Sensemaking Online: A Content Analysis of Facebook Posts by COVID-19 Long Haulers | Community and Support |

In emerging interdisciplinary research areas such as Long COVID, the capacity to categorize studies without manually curated labels represents a novel contribution. Fields involving public health, computational linguistics, and social sciences often converge on complex research questions, making any single classification system insufficient on its own. The zero-shot framework addressed this issue by enabling rapid, yet reliable, placement of studies within relevant categories, facilitating a coherent view of how different facets - like symptom trajectories, policy guidance, and community engagements - interact in the evolving literature. Such a methodology also offers a blueprint for future works that require the integration of heterogeneous sources of knowledge, allowing researchers to devote more time to interpreting outcomes rather than refining labeling procedures. In Section 4, a review of all these studies is presented.

## 4 Review of Papers

In this section, each study that met the inclusion criteria of this work has been reviewed under one of four broad areas, according to assignments generated by the zero-shot learning model (discussed in Section 3). This automated classification process distinguished primary thematic emphases among the papers, placing them into "NLP and Modeling," "Policy and Advocacy," "Community and Support," or "Symptom Characterization." The subsequent sections explore each area, discussing how individual studies addressed multimodal forms of social media-based inquiries into Long COVID.

### 4.1 NLP and Modeling

Fu [42] studied concerns about Long COVID as expressed on social media. They analyzed 117,789 tweets from March 2022 to April 2022 and utilized sentiment analysis and topic modeling. Their objectives included identifying emergent themes from users' experiences, such as the social and economic burdens tied to Long COVID. They observed that negative attitudes toward Long COVID were especially widespread and noted that such sentiments raised important considerations for clinicians and policymakers. Jordan et al. [75] conducted an investigation that combined a mixed approach with topic modeling. They gathered online data from medical sources, news outlets, and self-identified "long haulers", highlighting how personal distress connected with dissatisfaction regarding the healthcare system in the context of Long COVID. They found multiple themes, most of which showed concerns related to Long COVID that had either been disregarded or insufficiently recognized.

Kusuma et al. [76] studied social media data to isolate the most frequently discussed topics and to identify influential users engaging with the concept of extended recovery in the context of Long COVID. They used social network analysis and sentiment analysis. They analyzed 119,185 tweets from 94,325 users to demonstrate how certain public figures or health professionals influenced these discussions. The findings of sentiment analysis showed that most of these tweets were negative. Thakur [77] studied 1,244,051 tweets about Long COVID with a specific focus on using VADER for sentiment analysis. The findings showed that the percentages of tweets with positive, negative, and neutral sentiments were 43.1%, 42.7%, and 14.2%, respectively. The findings of this study also showed that most tweets with a positive sentiment and most tweets with a negative sentiment were not highly polarized.

The study by Koss et al. [82] explored the feasibility of social media mining methods to extract insights shared by Long COVID patients. They focused on extracting insights from Reddit ("/r/covidlonghaulers"), where participants described supplements and medications that they tested for symptom relief. Using named-entity recognition, they mapped out networks to illustrate how certain substances - such as magnesium, vitamins, and steroids - appeared frequently and often in connection with each other on Reddit. Jordan [96] conducted text-mining of Long COVID-related content on YouTube. They collected transcripts and comments to learn how self-identified "long haulers" perceived their illnesses and how broader audiences reacted. Jordan identified recurring issues that spanned uncertainty regarding medical systems, misinformation, and the need for coping strategies by applying topic modeling. The work of Awoyemi et al. [78] involved another exploration of tweets about Long COVID. Their work's initial data mining process resulted in 62,232 tweets, which were reduced to 10,670 tweets after removing the duplicates. Their study showed that the majority of the tweets about Long COVID originated from the United States of America (38%), United Kingdom (30%), and Canada (13%), with the most common hashtags being #longcovid (36%) and #covid (6.36%), and the most frequently used word being people (1.05%). They also performed sentiment analysis, which showed that the top three emotions detected in these tweets were trust (11.68%), fear (11.26%), and sadness (9.76%).

### 4.2 Policy and Advocacy

Rushforth et al. [63] used narrative inquiry and analyzed a dataset of narrative interviews and focus groups with 114 people with Long Covid from the United Kingdom, drawing on socio-narratology, therapeutic emplotment, and polyphonia. Their study showed how these personal stories served as catalysts for policy efforts and structural reforms, emphasizing how influential firsthand accounts can be helpful in compelling decision-makers to address an emerging public health challenge such as Long COVID.

Bhattacharyya et al. [80] studied tweets about Long COVID to understand the need for more resources to investigate the extended trajectory of COVID-19. They used the National Research Council (NRC) Emotion Lexicon method for sentiment analysis and identified an association between retweets and favorite counts on Twitter and particular emotional reactions, such as sadness, joy, or trust. Laestadius et al. [85] examined how US state health departments used Facebook for public messaging about COVID-19, with a particular focus on mentions of Long COVID. Their study identified 49,310 pandemic-related posts, with fewer than 200 explicitly discussing Long COVID. Using quantitative content analysis methods, they coded these posts about Long COVID. The results showed that 75.18% included language about susceptibility, 64.96% severity, and 64.23% benefits of prevention. In addition to this, cues to preventive action appeared in 54.01% of posts and 19.71% of posts provided guidance for those with Long COVID. Déom et al. [99] used a mixed-methods approach to analyze policy documentation and social media discourse about children and teenagers suffering from Long COVID in the United Kingdom. The authors used the LISTEN framework to demonstrate inconsistency in how guidelines reached the public and to emphasize the demand for mental health services for children, young people, and healthcare workers suffering from Long COVID. In their work, they also presented several policy recommendations, such as enhancing accountability through regular audits, promoting inclusiveness by incorporating perspectives of children and young people, ensuring transparency via regular updates, and maintaining equity in policy impact.

### 4.3 Community and Support

Russell et al. [64] investigated how online communities offered solace and informational resources to individuals with Long COVID symptoms. Through qualitative interviews, they found that people experiencing these symptoms went through significant ambiguity, which was often made worse by invalidation or denial in clinical settings. The findings of their work showed that online communities filled this gap by offering mutual support and reassurance, thus illustrating the essential psychological role that social networks can serve. Meledandri [65] performed a quantitative and qualitative evaluation of approximately 600,000 tweets about Long COVID. Their study showed that some of these tweets reflected conspiracy theories involving vaccination, fake news, and post-truths, clashing with scientific evidence, and the remaining tweets reflected supportive stances. Rhodehamel [79] studied the public discourse about Long COVID on Reddit (r/covidlonghaulers). Their study showed that community in the context of Long COVID was built on Reddit through three main themes. First, hope through validation, knowledge sharing, and helpfulness. Second, kinship through commiseration and shared experiences of suffering. Finally, the discourse surrounding harm, including ableism, grifting, exploitation, infighting, and tensions between people suffering from Long COVID and society.

Garrett et al. [84] investigated the experiences of women with Long COVID symptoms with a specific focus on how social media played a dual role in either nurturing or undermining well-being. The study showed that the main roles of social media included facilitating support group participation, experience sharing, interpersonal connections, and media consumption. The study also showed that participants relied on social media to fulfill their emotional support, social engagement, spirituality, health planning, information gathering, professional support, and recreational relaxation needs. The work done by Santarossa et al. [87] aimed to investigate the #longCOVID and #longhaulers conversations on Twitter using topic modeling and social network analysis. The findings of their work showed that among the 2010 tweets about long COVID-19 and 490 tweets by COVID-19 long haulers, 30,923 and 7817 unique words were found, respectively. Their work also showed that for both conversation types, "#longcovid" and "covid" were the most frequently mentioned words, and words relevant to having Long COVID were more frequently found in tweets posted by individuals suffering from Long COVID.

Miyake et al. [94] studied social media data collected at different points of the pandemic to explore how patients felt when official communications diverged from their lived experiences. They used a mixed methods approach involving quantitative and qualitative analyses and studied 1.38 million posts about Long COVID from Twitter, Facebook, blogs, and forums. The results indicated that the negative impacts arise mostly from conflicting definitions of COVID-19 and fears around the COVID-19 vaccine for individuals suffering from Long COVID. Their study also identified that key areas of concern in the context of Long COVID included time or duration, symptoms or testing, emotional impact, lack of support, and resources. Juanals et al. [97] studied Long Covid coverage by the French press and Twitter. More specifically, the objectives of their study were to analyze the modalities of construction and progressive visibility of Long Covid in the public media and on Twitter and to propose a methodology based on topic modeling and related concepts in NLP to conduct a comparative analysis between newspapers and Twitter coverage of Long COVID.

Ozduran et al. [98] classified YouTube videos about Long COVID according to video parameters and content analysis. They also determined the quality, reliability, and accuracy of these videos using the Global Quality Score (GQS), the Journal of American Medical Association (JAMA) Benchmark Criteria, and the Modified DISCERN Questionnaire. The findings showed that out of 180 videos about Long COVID, 74 were of low quality, 14 were of moderate quality, and 12 were of high quality; 21% contained insufficient data, 73% contained partially sufficient data, and 6% contained completely sufficient data. Their work also showed that videos uploaded by academic sources (66.7%) and physicians (12.5%) made up most of the high-quality group. The authors also found a statistically significant correlation between the source of upload and the number of views (p = 0.014), likes (p = 0.030), comments (p = 0.007), and video duration (p = 0.004).

### 4.4 Symptom Characterization

Singh et al. [66] focused on identifying symptoms on Twitter where users self-reported Long COVID. They studied the tweets published by 89 Twitter users, and the findings of their study showed that most users described multiple symptoms, out of which the most common were fatigue, shortness of breath, pain, and brain fog or concentration difficulties. Ziauddeen et al. [67] conducted an online survey with 2,550 participants, to highlight the range of symptoms of Long COVID and infer how such symptoms affected daily functioning. Their study showed that most participants described fluctuating (57.7%) or relapsing symptoms (17.6%), with physical activity, stress, and sleep disturbance being the commonly triggered symptoms. Their study also found that one-third of participants reported being unable to live alone without assistance six weeks from the start of the illness, and 16.9% reported being unable to work alone due to COVID-19 illness. The goal of the work done by Sarker et al. [68] was to infer Long COVID symptoms self-reported by users, compare symptom distributions across studies, and create a symptom lexicon by studying Long COVID-related posts on Reddit. They studied 42,995 posts by 4249 Reddit users, and the results showed that 1744 users expressed at least one symptom. The results of their work also showed that the most frequently reported long-COVID symptoms were mental health-related symptoms (55.2%), fatigue (51.2%), general ache or pain (48.4%), brain fog or confusion (32.8%) and dyspnea (28.9%). Banda et al. [69] used a combination of machine learning, natural language processing techniques, and clinician reviews and mined 296,154 tweets about Long COVID. The objective of their study was to characterize the course of Long COVID, create detailed timelines of symptoms and conditions, and analyze their symptomatology during a period of over 150 days.

Massey et al. [70] posted a poll to a Facebook group of 169,900 members that asked about breakthrough COVID-19 cases, Long Covid, and hospitalizations. The findings showed that out of the 1,949 participants who responded to the poll, 44 reported a symptomatic breakthrough case, and 24 reported that COVID-19 led to symptoms of Long COVID. Their study also found that 1 out of these 24 cases was hospitalized. The goal of the research by Martin et al. [71] was to explore healthcare workers' perceptions concerning Long COVID in children and young people in the UK between January 2021 and January 2022 by studying relevant posts on Twitter. This research showed that healthcare workers were responsive to announcements issued by authorities regarding the management of COVID-19 in the UK, and the most frequent emotion expressed on Twitter in this regard was negative. This research also identified the main themes of conversation, which included uncertainty about the future, policies and regulations, managing and addressing COVID-19 and Long COVID in children and young people, vaccination, using Twitter to share scientific literature and management strategies, and clinical and personal experiences.

Dolatabadi et al. [72] aimed to determine the validity and effectiveness of advanced NLP approaches to derive insight into Long COVID-related patient-reported health outcomes from social media platforms. They used Transformer-based BERT models to extract and normalize long COVID symptoms and conditions from English posts on Twitter and Reddit. The results indicated that the top three most commonly occurring Long COVID symptoms were systemic (such as "fatigue"), neuropsychiatric (such as "anxiety" and "brain fog"), and respiratory (such as "shortness of breath").

Miao et al. [73] used an interactive information extraction tool and analyzed tweets about Long COVID. The authors extracted key information from the relevant tweets and analyzed the user-reported Long COVID symptoms concerning their demographic and geographical characteristics. Feng et al. [74] also studied tweets about Long COVID. They classified Long COVID-related tweets into four categories based on the content, detected the presence of six basic emotions, and extracted prevalent topics. Their analyses revealed that negative emotions dominated throughout the study period.

Matharaarachchi et al. [81] implemented association rule mining to understand the patterns and behavior of long COVID symptoms reported by patients on Twitter. They found that in the 30,327 tweets included in their study, the most frequent symptoms were brain fog, fatigue, breathing or lung issues, heart issues, flu symptoms, depression, and general pain; loss of smell and taste, cold, cough, chest pain, fever, headache, and arm pain were noted in 1.6% to 5.3% of patients with long COVID. Ayadi et al. [83] collected 27,216 Reddit posts about Long COVID and performed a comprehensive data analysis. They found that over 78% of the analyzed posts referenced at least one symptom of Long COVID. The most reported symptoms were fatigue (29.4%), pain (22%), brain fog (19.1%), anxiety (17.7%), and headaches (15.6%). These symptoms frequently co-occurred with others, such as fever and nasal congestion. The symptoms were categorized into general (45.5%), neurological (42.9%), mental health or psychological or behavioral (35.2%), body pain or mobility (35.1%), and cardiorespiratory (31.2%).

Izquierdo-Condoy et al. [86] conducted a cross-sectional analysis of 2,103 participants in Ecuador between April and July 2022, using an online self-reporting questionnaire to investigate Long COVID symptoms. Among the respondents, 52.3% (1,100) reported Long COVID symptoms, with the majority being women (64%) and individuals aged 21-40 years (68.5%). Notably, 71.7% of the Long COVID cases occurred among residents at high altitudes (>2500m), compared to 29.3% at lower altitudes. Common symptoms included fatigue (8.4%), hair loss (5.1%), and difficulty concentrating (5.0%). The study identified a greater prevalence of Long COVID symptoms among women, individuals with severe initial infections, and those with comorbidities, emphasizing the influence of altitude on Long COVID. Déguilhem et al. [88] conducted a comprehensive analysis of 15,364 messages from 6,494 individuals with Long COVID or their caregivers in France from January 1, 2020, to August 10, 2021. The study identified three primary symptom co-occurrences: asthenia-dyspnea (35.3%), asthenia-anxiety (22.5%), and asthenia-headaches (17.3%). Key difficulties reported by patients included managing symptoms (35.4% of messages), dealing with psychological impacts such as anxiety and uncertainty (15.1%), enduring pain (12.0%), and coping with disruptions (9.4%) and professional life (8.0%). The analysis also categorized patients into three distinct profiles. Profile A consisted of 406 patients who exclusively reported asthenia. Profile B included 129 patients who predominantly experienced anxiety (100%), along with asthenia (21.7%), dyspnea (11.6%), and ageusia (2.3%). Profile C, with 141 patients, was characterized by dyspnea (100%) and asthenia (31.9%).

Additionally, the findings revealed that 49.1% of users expressed symptoms beyond three months post-infection, and 20.5% continued to report symptoms even after one year.

Segneri et al. [91] analyzed the communication style and network structure of 6,107 Reddit users to identify social traits associated with Long COVID. The study categorized users into three groups: No COVID (2,529 users), COVID (592 users), and Long COVID (2,986 users). They analyzed pre-pandemic posts, totaling 984,625, with 45% from the No COVID group, 32% from the Long COVID group, and 23% from the COVID group. Key findings from their work indicated that Long COVID users exhibited lower social media activity and fewer connections than other groups. Furthermore, their communication style included more health-related topics and frequent use of first-person singular pronouns but fewer anger-related words. Their study also found that Long COVID users were more likely to use interrogative language and verbose posts, reflecting their focus on health concerns. Sarker [93] studied self-reported Long COVID symptoms on the subreddit /r/covidlonghaulers. Using natural language processing, they identified the most common symptoms, including anxiety or stress, fatigue, body pain, and brain fog. Their study showed that most users reported 1-5 symptoms, with a median of 4 symptoms per user. Jacques et al. [100] analyzed the 100 most-viewed YouTube videos discussing Long COVID symptoms, uploaded between July 2020 and December 2021, which amassed 15,319,997 views. They found that the majority of these videos originated from television or internet-based news sources (56%), followed by consumer-generated content (32%), health professionals (9%), and entertainment TV (3%). Their study inferred that physical symptoms such as fatigue (73%), difficulty breathing (56%), and joint or muscle pain (49%) were most frequently discussed, alongside cognitive issues like brain fog (69%). Other frequently reported challenges included worsening symptoms post-activity (37%) and psychological effects like anxiety or depression (17%). Their work also showed that videos from entertainment TV received significantly more likes than other categories.

Strain et al. [101] surveyed 812 individuals with Long COVID to evaluate the impact of COVID-19 vaccination on their symptoms. The participants, primarily younger females (80.6%), reported symptoms persisting for over nine months in 71.6% of cases. Following the first vaccination dose, 57.9% of participants reported overall symptom improvement, while 17.9% experienced deterioration and the rest reported no change. Improvements were more pronounced with mRNA vaccines, such as Moderna (31% improvement) and Pfizer (24.4%), compared to the AstraZeneca adenoviral vector vaccine (22.6%). The most improved symptoms included fatigue ($p = 0.009$), brain fog ($p = 0.01$), and myalgia ($p = 0.006$). Their work found that symptom severity reductions were proportional to baseline scores and for half of the participants, symptom improvements were temporary, lasting 14-21 days, while post-vaccination deterioration resolved within 3-7 days.

The breadth of research surveyed in these four categories underscores how multifaceted Long COVID can be, spanning elements of advanced computational analyses, policy formation, community engagement, and clinical symptom documentation. Whether investigators used sentiment analysis to map public anxieties, examined official health communications to reveal messaging gaps, or monitored online forums where individuals assembled in search of guidance, each study contributed a piece to the broader puzzle of persistent post-COVID complications. In a collective manner, these works indicate that fully understanding Long COVID requires both interdisciplinary collaboration and innovative methodologies, particularly as social media continues to serve as a large-scale repository of patient-driven experiences. While some researchers focused on capturing emergent narratives via topic modeling, others underscored the imperative of supporting those living with Long COVID. By merging perspectives from computational sciences, public health, and real-world patient accounts, this body of literature highlights the need for sustained inquiry into Long COVID and for responsive frameworks that can adapt to new findings as they emerge.

## 5    Research Gaps and Future Directions

Although numerous investigations have emerged around Long COVID and its social media narratives, critical gaps warrant further consideration. One prominent challenge lies in harmonizing definitions and frameworks for the condition itself. Several studies used different inclusion criteria, with some focusing on self-reported experiences and others requiring clinical diagnoses. This lack of uniform standards hampers efforts to compare outcomes across different populations and time periods. Equally important is integrating finer details on patients' backgrounds, disease histories, and underlying conditions into study designs. Greater clarity in such baseline data could highlight whether certain cohorts, including older adults, are more prone to persistent complications [107-110]. In doing so, investigators may infer how aging-related factors, such as immunosenescence or latent comorbidities, interact with ongoing COVID-19 symptoms [111,112].

A second research gap concerns the depth and breadth of symptom documentation. While investigators have cataloged a wide array of complaints - ranging from cognitive difficulties to cardiorespiratory issues - there is still a shortage of longitudinal data that detail how and why certain symptoms linger or transform over time. More robust prospective studies could help pinpoint potential transition points or flare-ups that individuals frequently mention in online communities. This endeavor would be particularly meaningful for older demographics, given that parallel research on aging has demonstrated the value of tracking gradual physiological changes across extended intervals [113-116]. Building on this approach, future work might measure how chronic inflammation or age-associated immune variations could shape Long COVID trajectories [117-119]. Such insights would not only clarify the causes of persistent symptoms but might also inform targeted interventions tailored to different life stages.

In addition to these gaps in symptom tracking, a more systematic exploration of social determinants of health is necessary. Many of the reviewed studies used social media posts without consistently capturing users' socioeconomic status, geographic context, or access to healthcare resources. Understanding how stressors - like insufficient medical support or economic hardship - can amplify Long COVID complications would be invaluable for refining health policies. It would also broaden our understanding of how aging adults, who may already be dealing with multiple conditions, manage additional burdens imposed by COVID-19 [120-123]. In this sense, bridging findings from gerontological research - where socioeconomic disparities often exacerbate the severity of age-related disorders - could enrich the framework for studying Long COVID's psychosocial dimensions.

Another significant avenue for future work involves methodological innovations, particularly those that incorporate advanced analytics beyond sentiment analysis or basic topic modeling. For example, multi-modal data analysis - integrating text, voice, and video - could provide a richer characterization of individuals' lived experiences with Long COVID. Such approaches might benefit from cutting-edge computational tools used in aging research, where sensors and wearables have been employed to track physiological and behavioral indicators of decline [124-126]. Translating these methods to the COVID-19 context could enable real-time monitoring of symptom fluctuations or early warning signs. Coordinated collaborations between specialists in gerontology and emerging fields like machine learning could foster the exchange of techniques that have proven effective in monitoring complex, chronic conditions [127-129].

Furthermore, many researchers have drawn attention to the lack of formal clinical trials or intervention studies aimed at mitigating long-term symptoms despite ample anecdotal evidence shared on social media. Addressing this shortfall requires not just larger sample sizes but also ethically designed studies that compare different management strategies - pharmacological, rehabilitative, or psychosocial. Such work would be especially beneficial if it includes older adults, given that gerontology has a long-standing tradition of rigorously testing interventions aimed at prolonging functional autonomy [130-133]. By synthesizing expertise from both Long COVID and aging-focused investigations, scholars could devise clinical protocols that emphasize the realities of multi-morbidity, polypharmacy, and overall resilience [134-136]. In effect, the field could progress toward interventions that systematically address both the biological underpinnings of prolonged COVID-19 symptoms and the socio-emotional hurdles encountered by diverse patient groups, including older populations.

These directions underscore the need for broader interdisciplinary engagement, deeper longitudinal insights, and more nuanced epidemiological tools. They also highlight how lessons from aging research can advance Long COVID research, especially regarding risk assessment, symptom evolution, and care strategies [135-140]. Future work could also focus on refining how social media data are aggregated and annotated. Researchers often concentrate on static posts [141,142], yet new formats, such as ephemeral stories [143,144] or live chats [145,146], offer dynamic insights into how users articulate their symptoms and needs in real-time. A structured, time-sensitive approach to analyzing these data would help identify swift changes in public sentiment or emerging topics that might otherwise go unnoticed. Complementary initiatives to standardize metadata collection could lay the groundwork for more consistent data sharing across research groups, encouraging broader comparative efforts and mitigating gaps in knowledge [147-150].

A related priority involves exploring how individuals engage with one another on social media beyond simple "like" or "share" metrics in the context of Long COVID-related discussions. In many online communities, participants build deep interpersonal networks that thrive on mutual trust, and these environments can shape patterns of symptom reporting and health-seeking behaviors [151,152]. Future studies could investigate how trust relationships form and evolve within these networks and how different age groups interpret and disseminate healthcare information online. Such inquiries could highlight how older adults, who may be managing multiple comorbidities, adapt to social media for guidance and peer support in ways that diverge from younger demographics. Another avenue involves enhancing the accuracy of sentiment and topic detection by adopting more context-aware algorithms. While many current models focus on keyword frequency or text structure, there is room for approaches capable of capturing subtler emotional or cultural nuances. Future works could consider leveraging contextual embeddings that adapt to evolving language trends, including new phrases or slang coined by social media communities [153-155]. By calibrating these advanced tools to different platforms - whether Reddit, Twitter, or localized forums - investigators could develop a finer-grained picture of how users articulate the long-term effects of COVID-19 in different social media platforms.

Researchers in this field could also explore closer collaborations with social media platforms themselves. Data-access initiatives and robust privacy safeguards might foster the co-design of tools and dashboards that allow public health professionals and community leaders to monitor and interpret conversations as they unfold. Such partnerships could promote real-time feedback loops, where findings from social media analysis guide new research questions and public health messaging, and vice versa. Over the long term, these initiatives may pave the way for better resource allocation and more precisely tailored interventions, ensuring that people suffering from Long COVID have reliable information and consistent support.

The work presented in this paper has a couple of limitations. The heterogeneous nature of social media platforms and varying user demographics and data availability may introduce sampling biases. Additionally, the zero-shot classification method, though novel, relies on pre-trained language representations and may not fully capture contextual nuances in highly specialized or region-specific vocabulary as expressed on social media.

## 6     Conclusion

Long COVID is becoming a complex health-related problem that requires multiple approaches to comprehensively study the wide range of clinical, long-term effects, and psychosocial aspects associated with it across patients worldwide. This review highlights the significant contribution of social media networks in furthering the understanding of Long COVID, depicting many facets of how patients experience the condition, the symptoms that typically develop, and how the general population understands this condition. Social media has allowed for the collection of patient-generated data in real-time, making it easier to represent the variety of symptoms associated with Long COVID. By presenting a systematic review of studies that rely on user-generated data on social media platforms and by using a transformer-based zero-shot learning approach, this paper offers new perspectives on how we can capture and categorize this complex research landscape. The review shows that patients' online narratives do far more than supplement clinical findings; they also create a real-time feedback loop that may guide research questions and shape both public health policies and advocacy efforts. These accounts often raise critical questions about persistent symptoms, gaps in healthcare, and the psychological burden of extended illness. By examining these issues within a coherent, data-driven framework, the paper shows how computational methods can highlight patterns and themes crucial for this research area.

However, the significance of this research goes beyond mapping out social media trends and insights. The zero-shot classification pipeline demonstrates that advanced

20language models can identify meaningful categories within research papers on emerging conditions, all while bypassing conventional manual labeling workflows. This streamlined process broadens the scope of discovery, allowing for timely insights in a field that demands ongoing updates. The insights presented here confirm a vital need for refining longitudinal approaches, standardizing both clinical and analytical frameworks, and exploring how diverse populations experience Long COVID. The future directions presented in this paper highlight that interdisciplinary research should utilize advanced computational tools that capture the nuanced language and behavior of social media users, especially as public discourse and scientific understanding related to Long COVID continues to evolve. Furthermore, investigators may also delve into multi-modal social media data - such as images, videos, and real-time audio streams - to supplement textual content. In summary, this paper's findings underscore the potential of combining patient-driven data with advanced analytics to refine our understanding of Long COVID, while also laying the groundwork for future work in this area that adapts to new discoveries and patient needs.